# *In situ* coherent diffractive imaging

Yuan Hung Lo[1,2], Lingrong Zhao[1,3], Marcus Gallagher-Jones[1], Arjun Rana[1], Jared Lodico[1], Weikun Xiao[2], B. C. Regan[1] and Jianwei Miao[1]

*[1]Department of Physics and Astronomy, and California NanoSystems Institute, University of California, Los Angeles, CA 90095, USA. [2]Department of Bioengineering, University of California, Los Angeles, CA 90095, USA. [3]Department of Physics and Astronomy, Shanghai Jiao Tong University, Shanghai 200240, China.*

**Coherent diffractive imaging (CDI) has been widely applied in the physical and biological sciences using synchrotron radiation, XFELs, high harmonic generation, electrons and optical lasers. One of CDI's important applications is to probe dynamic phenomena with high spatio-temporal resolution. Here, we report the development of a general *in situ* CDI method for real-time imaging of dynamic processes in solution. By introducing a time-invariant overlapping region as a real-space constraint, we show that *in situ* CDI can *simultaneously* reconstruct a time series of the complex exit wave of dynamic processes with robust and fast convergence. We validate this method using numerical simulations with coherent X-rays and performing experiments on a materials science and a biological specimen in solution with an optical laser. Our numerical simulations further indicate that *in situ* CDI can potentially reduce the radiation dose by more than an order of magnitude relative to conventional CDI. As coherent X-rays are under rapid development worldwide, we expect *in situ* CDI could be applied to probe dynamic phenomena ranging from electrochemistry, structural phase transitions, charge transfer, transport, crystal nucleation, melting and fluid dynamics to biological imaging.**

The first experimental demonstration of CDI in 1999[1] has spawned a wealth of development in lensless imaging and computational microscopy methods with widespread scientific applications[2-32]. With continuous rapid development of coherent X-ray sources[33-36], high-speed detectors[37] and powerful algorithms[38,39], CDI methods are expected to have a larger impact across different disciplines in the future[36]. As many natural phenomena of interest evolve in response to external stimuli, CDI can make important contributions to the understanding of these dynamical phenomena[22,29,36,41,42]. Recently, *in situ* and *operando* X-ray microscopy have advanced rapidly to study dynamic processes with elemental and chemical specificity[43,44], but the spatial resolution is limited by the X-ray lens. While *in situ* electron microscopy can achieve much higher spatial resolution[45], the dynamic scattering effect limits the sample thickness and restricts the technique's applicability to a wider range of samples. In this article, we demonstrate a general *in situ* CDI method to simultaneously reconstruct time-evolving complex exit waves of dynamic processes with spatial resolution only limited by diffraction signals. By introducing both static and dynamic regions in the experimental geometry, we apply the static region as a powerful time-invariant constraint to reconstruct the dynamic process of extended samples with fast and robust convergence. Our numerical simulations indicate that with advanced synchrotron radiation, *in situ* CDI could potentially achieve 10 nm spatial resolution and 10 ms temporal resolution. Using an optical laser, we conduct proof-of-principle experiments of this method by capturing the growth of Pb dendrites on Pt electrodes immersed in an aqueous solution of $Pb(NO_3)_2$ and by reconstructing a time series of phase images of live glioblastoma cells in culture medium. Furthermore, by varying the incident X-ray flux between the static and dynamic regions, we demonstrate through numerical simulations that *in situ* CDI can potentially reduce the radiation dose to radiation sensitive samples by more than an order of magnitude relative to conventional CDI.

### *In situ* CDI principle

To achieve fast, reliable reconstruction of a time series of dynamic phenomena, *in situ* CDI takes advantage of two types of structures or regions. A dynamic region constantly changes over time or in response to external stimuli, while a static region remains stationary in time. A time series of far-field diffraction patterns are collected with interference between the static and dynamic regions. Since the static region remains unchanged during the data acquisition, this interference effectively creates a time-invariant overlapping region between the measured diffraction patterns, providing a powerful real space constraint to simultaneously phase all diffraction patterns with fast and robust convergence. Figure 1a shows an experimental setup for *in situ* CDI. A dual-pinhole aperture is placed upstream of the sample to create two separate regions on the sample plane. The dynamic specimen of interest is localized to the area of one pinhole, while the other pinhole illuminates a region without the sample. Note that the second, static region can be completely empty or a substrate containing some stationary structure. Experimentally, the sample holder can be prepared by using microfluidics so that there are regions where one pinhole occupies the dynamic specimen while the other one covers a static area (Fig. 1a). Furthermore, this technique can be used to extend scanning CDI techniques such as ptychography, where a region of interest can first be obtained by scanning, and then the dynamic specimen can be magnified and perturbed to probe dynamic information. As a general method, *in situ* CDI requires only a static region or structure between two consecutive time frames as the time-invariant constraint for phase retrieval, which can in principle be implemented with different experimental geometries.

### *In situ* CDI phase retrieval algorithm

Figure 1b shows the schematic layout of the *in situ* CDI phase retrieval algorithm. Using random phases as an initial input, the algorithm iterates between real and reciprocal space with constraints incorporated in each space. The illumination function of the incident wave and a static function of the time-invariant overlapping region are enforced in real space, while the measured Fourier magnitudes are applied in reciprocal space. In each iteration, a weighted average static function is sequentially passed onto the reconstructions of the time series. Since the static function is shared and mutually reconstructed at different time frames, the solutions to the phase problem for the whole time series rapidly emerge without stagnation. The $j^{\text{th}}$ iteration of the algorithm consists of the following steps.

i) Obtain a weighted average static function at time $t$

$$S'_{t,j}(\boldsymbol{r}) = \gamma S_{t-1,j}(\boldsymbol{r}) + (1-\gamma)S_{t,j}(\boldsymbol{r}), \qquad (1)$$

where $S_{t-1,j}(\boldsymbol{r})$ represents the static function at time $t$-1 and the weighting factor $\gamma$ is set to 0.8.

ii) Combine $S'_{t,j}(\boldsymbol{r})$ with a dynamic function, $D_{t,j}(\boldsymbol{r})$, to produce an object function

$$O_{t,j}(\boldsymbol{r}) = S'_{t,j}(\boldsymbol{r}) + D_{t,j}(\boldsymbol{r}). \qquad (2)$$

iii) Multiply the object function by the illumination function, $P(\boldsymbol{r})$, to generate a complex exit wave function

$$\psi_{t,j}(\boldsymbol{r}) = O_{t,j}(\boldsymbol{r})P(\boldsymbol{r}). \qquad (3)$$

In the current version of the algorithm, an accurate knowledge of $P(\boldsymbol{r})$ is necessary, and can be experimentally measured[6,11,46].

iv)    Apply the fast Fourier transform (FFT) ($\mathcal{F}$) to the exit wave function to obtain its Fourier transform

$$\Psi_{t,j}(\boldsymbol{k}) = \mathcal{F}\big[\psi_{t,j}(\boldsymbol{r})\big]. \quad (4)$$

v)    Replace the calculated Fourier magnitude with the measured one

$$\Psi'_{t,j}(\boldsymbol{k}) = |\Psi^m_t(\boldsymbol{k})| \, \frac{\Psi_{t,j}(\boldsymbol{k})}{|\Psi_{t,j}(\boldsymbol{k})|}. \quad (5)$$

vi)    Apply the inverse FFT ($\mathcal{F}^{-1}$) to obtain an updated exit wave function

$$\psi'_{t,j}(\boldsymbol{r}) = \mathcal{F}^{-1}[\Psi'_{t,j}(\boldsymbol{k})]. \quad (6)$$

vii)    Remove $P(\boldsymbol{r})$ to get an updated object function

$$O'_{t,j}(\boldsymbol{r}) = O_{t,j}(\boldsymbol{r}) + \frac{|P(r)|P^*(r)}{\alpha(|P(r)|^2+\varepsilon)}\big[\psi'_{t,j}(\boldsymbol{r}) - \psi_{t,j}(\boldsymbol{r})\big], \quad (7)$$

where $\alpha = max|P(r)|$ and $\varepsilon$ is a small value to prevent division by 0 (ref. 46).

viii)    Separate $O'_{t,j}(\boldsymbol{r})$ into the updated static and a dynamic functions, $S_{t+1,j}(\boldsymbol{r})$ and $D_{t,j+1}(\boldsymbol{r})$, respectively, and feed $S_{t+1,j}(\boldsymbol{r})$ back to step i) to reconstruct $D_{t+1,j}(\boldsymbol{r})$.

ix)    After repeating steps i)-viii) for the whole time series, an $R$-factor is calculated for the $j^{th}$ iteration to monitor the convergence of the algorithm

$$R_j = \frac{\sum_t \sum_k \big||\Psi^m_t(\boldsymbol{k})| - |\Psi_{t,j}(\boldsymbol{k})|\big|}{\sum_t \sum_k |\Psi^m_t(\boldsymbol{k})|}. \quad (8)$$

After several hundred iterations, the algorithm quickly converges to the correct solution even in the presence of noise and missing data. Another unique feature of the algorithm is its ability to simultaneously reconstruct the complex exit waves of all frames without the necessity of averaging independent runs for individual frames.

**Numerical simulations of *in situ* CDI using coherent X-rays**

Batteries play an indispensable role in the development of modern technologies, but advances in high capacity batteries are hampered by dendritic growth, where microfibers of electrolyte materials sprout from the surface of electrodes during charge/discharge cycles and short the circuit. In some serious cases dendrites can cause rapid heating and explosion of the battery[47]. While *in situ* TEM can observe dendritic growth at high spatial resolution[48], the sample thickness is limited by the dynamical electron scattering effect and the temporal resolution is hampered by the electron flux[45]. Due to X-ray's larger penetration depth, *in situ* CDI is ideally suited to probe the dynamic phenomena of thick specimens with nanoscale spatial resolution and high temporal resolution. To demonstrate *in situ* CDI's ability to reliably reconstruct dynamic structures, we performed numerical simulations on real time imaging of Pb dendrite growth in solution (Methods) (Fig. 2). Coherent X-rays with 8 keV energy and a flux of $10^{11}$ photons/$\mu m^2$/s were incident on a dual-pinhole aperture. The illumination function was generated by propagating an exit wave from the dual-pinhole aperture to the sample plane. One of the pinholes illuminated the growth process of Pb dendrites immersed in a 1-$\mu$m-thick water layer (Methods), while the other pinhole was focused on a static region. A time series of diffraction patterns were collected by a 1024×1024 pixel detector with a frame rate of 100 Hz and a linear oversampling ratio of ~2 (ref. 48). Poisson noise was added to each diffraction pattern and the central 5×5 pixel data was removed to simulate the missing center problem (Fig. 2a). By using random phase sets as an initial input, the *in situ* CDI algorithm quickly converged to the correct solution after several hundreds of iterations (Fig. 2b). Figure 2d shows a time series of the magnitudes of the reconstructed complex exit waves with a temporal resolution of 10 ms, which are in good agreement with the original structure model (Fig. 2c). Compared to conventional phase retrieval algorithms[38,39,50-52], the *in situ* CDI algorithm can produce very consistent final reconstructions with different random phase sets as the initial input. To quantify the reconstructions, we calculated the Fourier ring correlation (FRC, Methods) between the reconstructed images and the original structure models, indicating a spatial resolution of 10 nm was achieved in this case (Fig. 2b).

**_In situ_ CDI experiment of a materials science sample using an optical laser**

As a proof-of-principle experiment, we demonstrated *in situ* CDI for materials science applications by capturing the growth of Pb dendrites on Pt electrodes immersed in an aqueous solution of $Pb(NO_3)_2$. A HeNe laser was used as the coherent light source and illuminated a dual-pinhole aperture composed of two 100 µm holes spaced 100 µm apart edge-to-edge (Methods). An electrochemical cell was placed 400 µm downstream of the aperture. The cell was made from 50 µm diameter Pt wires immersed in 1.5M $Pb(NO_3)_2$ solution and encased between two 100-µm-thick coverslips (Methods). The left pinhole was placed in front of the electrochemical cell, while the right pinhole was focused on the substrate devoid of any dendrite. Twelve DC voltages were applied to the electrochemical cell to generate Pb dendrite growth and dissolution. At each voltage, a diffraction pattern was measured by a liquid-nitrogen cooled CCD detector with 1340×1300 pixels and a pixel size of 20×20 µm. To validate the *in situ* CDI results, a 5×5 ptychographic scan was also collected at each voltage. Supplementary Fig. 1, Figs. 3a and b and show the *in situ* CDI and ptychographic reconstructions of the same sample area at 12 different voltages. The overall structures are in good agreement between the two methods and the independent *in situ* CDI reconstructions are also very consistent (Supplemental Fig. 2). Figures 3c and d show some fine features are resolved in *in situ* CDI, but blurred in the ptychographic reconstruction. This blurring is due to continuous dendrite dissolution as the aperture scans over the field of view, resulting in an average reconstruction within the ptychographic scan.

Our results show that, as the voltage was ramped up to 1.8 V, Pb was rapidly deposited on the tip of the Pt wire to form short and wide dendrites (Supplementary Video 1). Initially the growth continued as the voltage decreased to 1.5 V, but as the potential decreased further the dendrite began to dissolve from its tip down to the root. The dendrite did not fully dissolve from the tip during the measurement, even after the voltage was reversed. The presence of undissolved Pb dendrites increases the surface roughness of the electrode and can lead to enhanced dendrite growth in subsequent charge/discharge cycles. This highlights dendrite growth as a significant problem in rechargeable batteries, where many repeated charge/discharge cycles occur over the lifespan of the battery[47].

### *In situ* CDI experiment of a biological sample using an optical laser

Tumor cell interaction offers insights into cancer progression, including recognition, communication and assembly among cell groups[53]. Tumor cell fusion, or fusogenic events, has also been suggested as a source of genetic instability, as well as mechanisms for metastasis and drug resistance[54]. The fate of fused cells could be either reproduction or apoptosis, with still unclear implications. To demonstrate *in situ* CDI in a biological context, we used a HeNe laser and collected a time series of 48 diffraction patterns from live glioblastoma cells sealed between two cover slips (Methods). To validate our method for imaging the biological specimen, a 5×5 ptychographic scan was also collected at each time point. Figure 4 and Supplementary Fig. 3 show a good agreement between the unwrapped phase images of *in situ* CDI and ptychographic reconstructions. The phase images show a small cell, about 25 μm in length, slowly approaching and attaching to a larger cell, about 100 μm in length, over two hours (Supplementary Video 2). After 144 minutes, the large cell responded to the presence of the smaller cell and underwent a rapid morphological change. In the next three hours the large cell moved away from the small cell as the small cell's thin pseudopodium anchored and pulled on the large cell to keep it in place. In the subsequent three hours the two cells fused together and formed a dense circular shape. Another time series of the cells taken after fusion showed no noticeable morphological change or cell motility, suggesting that apoptosis occurred after the cells merged.

**Numerical simulations on potential significant radiation dose reduction using *in situ* CDI**

To image radiation sensitive specimens with X-rays, the radiation damage process ultimately limits the achievable resolution[55,56]. One area currently being explored is the addition of a known diffusive structure to the sample to enhance the scattered signal. Placing a high-Z element structure in the field of view has demonstrated the possibility of reducing the dose required for obtaining minimum reconstructible diffraction signal[57-61]. Furthermore, since photons incident on the static region in *in situ* CDI do not hit the sample, those photons can enhance the measurable signal without inducing extra radiation damage to the sample. A carefully constructed static structure may also be used as additional *a priori* information to aid phase retrieval. Exploring a combination of these dose reduction strategies can help advance *in situ* CDI toward dynamic imaging of radiation sensitive samples. To examine this idea, we simulated a static structure of a 20-nm-thick Au pattern (Fig. 5a) and a biological sample consisting of a vesicle and protein aggregates (Fig. 5b, table 1, Methods). Both the static structure and biological sample are submerged in 1-μm-thick $H_2O$ and masked by a 3 μm pinhole. Using coherent soft X-rays (E = 530 eV), we first calculated the diffraction patterns only from the biological sample with a total fluence varying from $3.5 \times 10^4$ to $3.5 \times 10^7$ photons/μm$^2$, corresponding to a radiation dose ranging from $2.75 \times 10^3$ to $2.75 \times 10^6$ Gy, respectively (Methods). The diffraction patterns were collected by a detector with quantum efficiency of 80% and Poisson noise was added to the diffraction intensity (Fig. 5c). By using the oversampling smoothness (OSS) algorithm[52], we reconstructed the electron density of the biological sample from these noisy diffraction patterns (Figs. 5e-h). To quantify the spatial resolution, we calculated the Fourier ring correlation between the reconstructions and the model (Fig. 5m). Based on the 1/e criterion, we estimated the achieved resolution as a function of the total fluence.

Next, we calculated the diffraction patterns from a combination of the biological sample and the static structure. The total fluence on the biological sample varies from $3.5 \times 10^4$ to $3.5 \times 10^7$ photons/μm$^2$, while the total fluence on the static structure is fixed at $1.4 \times 10^{10}$ photons/μm$^2$. Experimentally, this can be implemented by introducing an absorber to the pinhole in front of the biological sample. The center-to-center distance between the biological sample and the static structure is 3.8 μm. Figure 5d shows the noisy diffraction pattern with a fluence of $3.5 \times 10^7$ photons/μm$^2$ on the sample and $1.4 \times 10^{10}$ photons/μm$^2$ on the static structure, which exhibits much spatial frequency diffraction signals than that calculated only from the biological sample with the same fluence (Fig. 5c). By using the static structure as a constraint, we reconstructed the electron density of the biological sample from the noisy diffraction patterns, showing significant improvement in image quality and spatial resolution (Figs. 5i-l). According to the Fourier ring correction (Figs. 5m and n), the *in situ* CDI method can reduce the radiation dose imparting on the sample by more than an order of magnitude, while maintaining the same spatial resolution. In some cases (e.g. $3.5 \times 10^4$ photons/$\mu m^2$ in Figs. 5m and n), the total dose can be reduced by two orders of magnitude with the same achievable resolution. Our numerical simulations suggest that the level of radiation dose reduction is related to the structure of the static pattern and the ratio of the coherent flux between the static and dynamic structure. To our knowledge, this could be the most dose efficient X-ray imaging method to probe radiation sensitive systems.

**Discussion**

*In situ* CDI overcomes a major challenge associated with traditional phase retrieval algorithms. In the presence of incomplete data and noise, conventional phase retrieval algorithms can be trapped in local minima and require averaging multiple independent runs to improve the final reconstruction[38,39,50-52]. By enforcing a time-invariant overlapping region as a powerful real-space constraint, *in situ* CDI is robust to incomplete data and noise and can simultaneously reconstruct a time series of complex exit waves without the stagnation issue or being trapped in local minima. The fine feature changes in the reconstructions between different time frames can be clearly distinguished. Furthermore, the experimental configuration of *in situ* CDI can be improved by using a dedicated (e.g. microfluidic) sample chamber (Fig. 1a), where the specimen of interest is physically separated from the static region by a barrier. Such customized experimental configuration could simplify the data collection and optimize the quality of reconstructions.

Compared to Fourier holography[62-64], *in situ* CDI has three unique distinctions. First, in Fourier holography, the spatial resolution is determined by the size of the reference source. In the X-ray regime, it is not only a challenge to fabricate very small reference sources, but also a small reference source would throw away a large fraction of coherent X-ray flux. On the other hand, *in situ* CDI does not have these limitations as its spatial resolution is only determined by the spatial frequency of the diffraction intensity. Second, Fourier holography calculates the autocorrelation function from the hologram using the inverse Fourier transform[62-64]. To extract the image of a sample from its autocorrelation function, the sample and the reference source must satisfy a geometry requirement. But *in situ* CDI uses an iterative algorithm for phase retrieval and has no geometry requirement between the static and dynamic structure. Third, in Fourier holography, the magnitude of the reference wave has to be comparable to that of the object wave for obtaining good quality autocorrelation functions. With *in situ* CDI, our numerical simulations indicate the coherent flux incident on the static and dynamic structure can vary by almost six orders of magnitude (Fig. 5i). Furthermore, by adjusting the coherent flux between the static and dynamic structure, one can potentially reduce the radiation dose to biological samples by more than an order of magnitude relative to conventional CDI (Fig. 5m).

**Conclusions**

We have developed a general *in situ* CDI method for simultaneously reconstructing a time series of complex exit waves of dynamic processes. We validate this method using both numerical simulations and experiments on materials science and biological samples. Our numerical results indicate that the combination of *in situ* CDI and advanced synchrotron radiation can be used to image dynamic processes in solution with a spatial resolution of 10 nm and a temporal resolution of 10 ms. Using an optical laser, we have performed proof-of-principle experiments of *in situ* CDI by capturing the growth of Pb dendrites on Pt electrodes immersed in an aqueous solution of $Pb(NO_3)_2$ and reconstructing a time series of the phase images of live glioblastoma cells in culture medium. There are four unique features associated with *in situ* CDI. First, it can simultaneously reconstruct a time series of complex exit waves with robust and fast convergence. Because no averaging is required in the reconstruction, fine structure variation at different time frames can be reliably reconstructed. Second, compared to liquid cell TEM[45], this method can be used to study the dynamics of a wider range of specimens (either thick or thin) in an ambient environment by optimizing X-ray energy based on the sample thickness and the chemical composition and reducing the multiple scattering effect. Third, while ptychography uses partially overlapping structure in the space domain as a constraint, *in situ* CDI uses partially overlapping structure in the time domain as a constraint. Furthermore, by avoiding the requirement of sample scanning, *in situ* CDI can achieve higher temporal resolution than ptychography. Finally, this *in situ* approach can be applicable to any type of radiation with flexible experimental geometry as long as a static structure can be used as a time-invariant constraint. The spatial and temporal resolution of the method is ultimately limited by the coherent flux and the read-out time of the detector. As coherent X-ray sources such as XFELs, advanced synchrotron radiation and high harmonic generation[33-36] as well as high-speed detectors[37] are under rapid development worldwide, we expect that this general *in situ*

CDI method can potentially open the door to imaging a wide range of dynamical phenomena with high spatio-temporal resolution.

**Acknowledgements** We thank D. K. Toomre for a stimulating discussion about the *in situ* CDI method. This work was supported by STROBE: A National Science Foundation Science & Technology Center under Grant No. DMR 1548924 and the DARPA PULSE program through a grant from AMRDEC.

Simulation geometry

| | |
|---|---|
| Detector size | 1100×1100 pixels |
| Detector quantum efficiency | 80% |
| Detector pixel size | 10 µm |
| X-ray energy | 530 eV |
| Sample-to-detector distance | 5 cm |
| Pinhole diameter | 3 µm |
| Fluence on dynamic structure | $3.5 \times 10^4 - 3.5 \times 10^7$ photons/µm$^2$ |
| Fluence on static structure | $1.4 \times 10^{10}$ photons/µm$^2$ |

Sample parameters

| | |
|---|---|
| Maximum dynamic structure thickness ($H_{50}C_{30}N_9O_{10}S_1$) | 1 µm |
| Static structure thickness (Au) | 20 nm |
| Protein density | 1.35 g/cm$^3$ |

**Table 1.** Dose reduction simulation parameters

**Figures and legends**

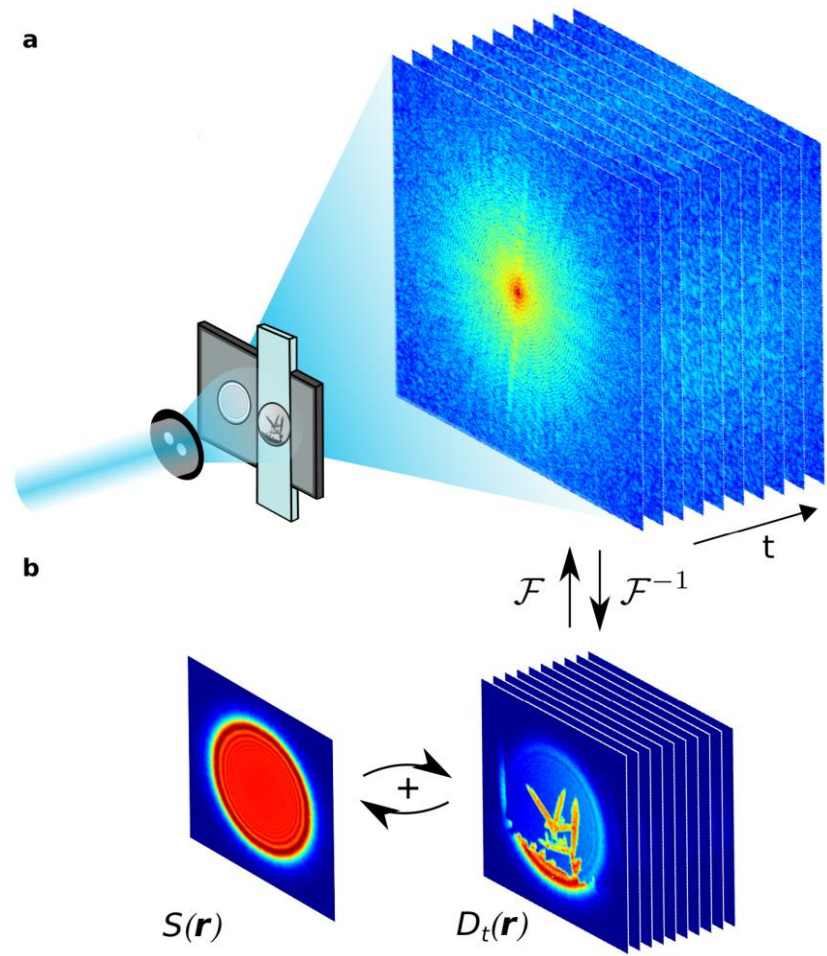

**Figure 1**. **Schematic layout of an experimental geometry and the phase retrieval of *in situ* CDI. a**, A coherent wave illuminates a dual-pinhole aperture to create a static and a dynamic region, $S(\boldsymbol{r})$ and $D_t(\boldsymbol{r})$. A sample in the dynamic region changes its structure over time and a time series of diffraction patterns are collected by a detector. **b**, By using the static region as a powerful time-invariant constraint in real space, the *in situ* CDI algorithm iterates between real and reciprocal space and simultaneously reconstructs a time series of complex exit waves of the dynamic processes in the sample with robust and fast convergence. $\mathcal{F}$ and $\mathcal{F}^{-1}$ represent the fast Fourier transform and its inverse.

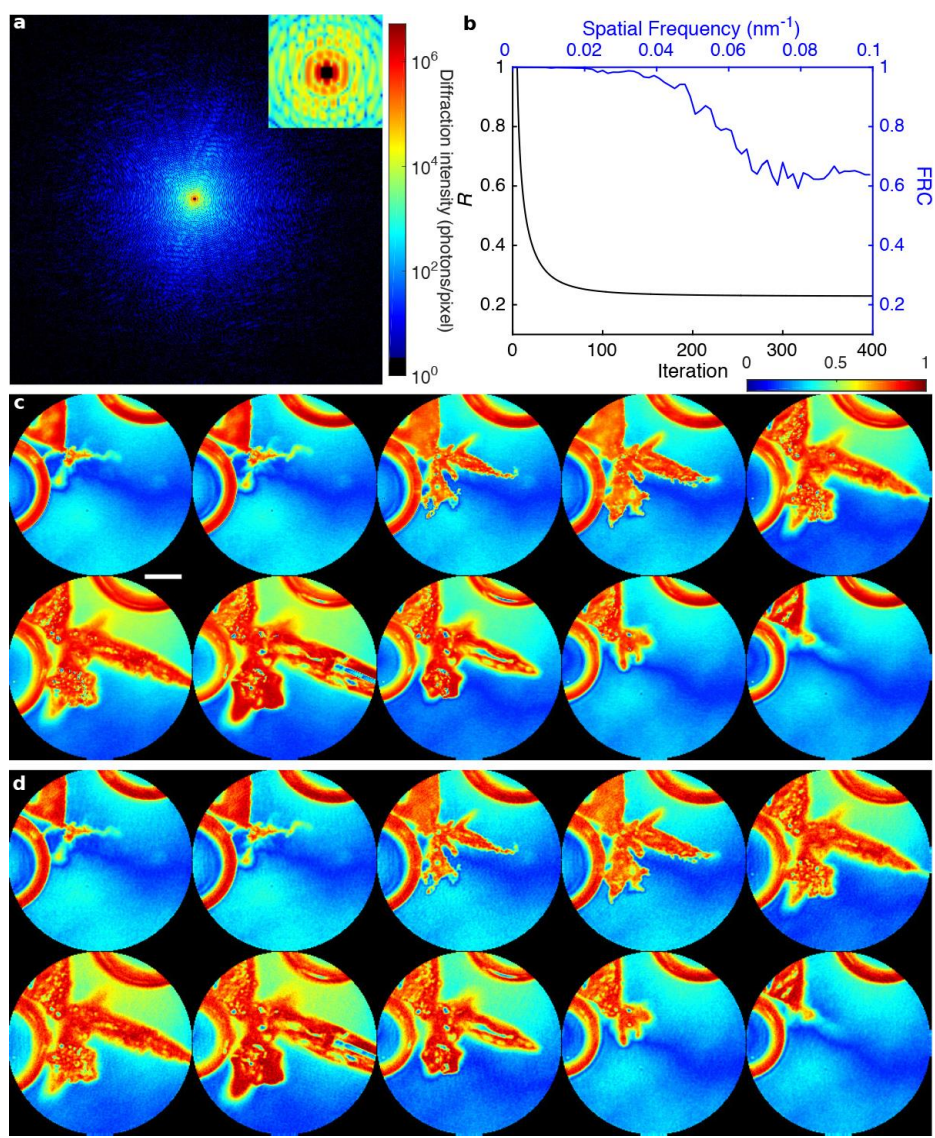

**Figure 2**. **Numerical simulations of *in situ* CDI with coherent X-rays. a**, A representative diffraction pattern with Poisson noise and missing data, calculated from the Pb dendrite formation process in an electrochemical cell using 8 keV X-rays with a flux of $10^{11}$ photons/$\mu$m$^2$/s. The insert indicates a 5x5 pixel missing data at the center. **b**, An R-factor (black curve) used to monitor the iterative algorithm, showing the rapid convergence of the algorithm. Average Fourier ring correlation (blue curve) between a time-evolving structure model and its corresponding reconstructions indicates a spatial resolution of 10 nm is achieved, with a temporal resolution of 10 ms. **c**, The time-evolving structure model of the dendrite formation process immersed in a 1-$\mu$m-thick water layer. Scale bar, 200 nm. **d**, The corresponding reconstructions of the time-evolving complex exit waves (showing only the magnitude), which are in good agreement with the structure model.

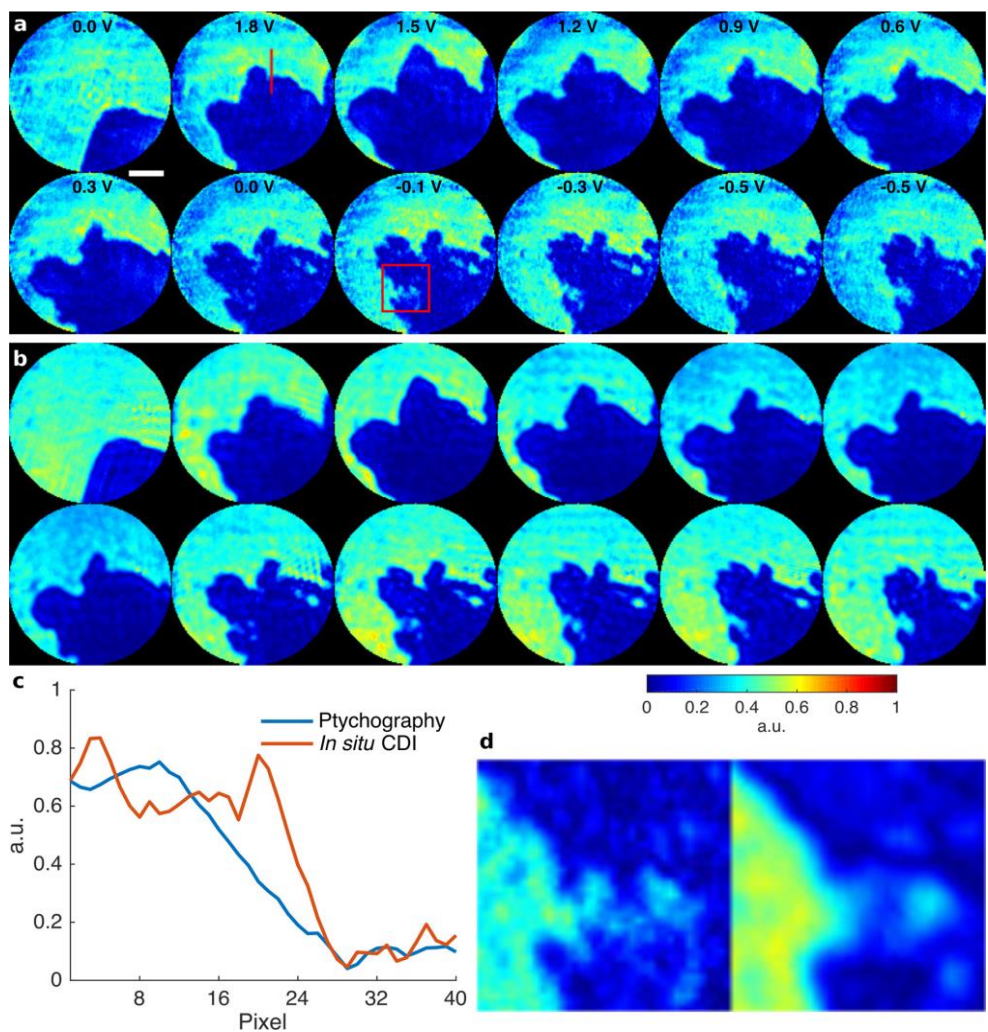

**Figure 3. Proof-of-principle experiment on *in situ* CDI with a materials science sample**. **a**, The magnitude of the complex exit waves reconstructed by *in situ* CDI, capturing the growth of Pb dendrites on Pt electrodes immersed in an aqueous solution of $Pb(NO_3)_2$ as a function of the applied voltage. Scale bar, 20 μm. **b**, Ptychographic reconstructions of the same dynamic sample area. The overall structures agree well between the two methods. However, some fine features are resolved in *in situ* CDI, but blurred in the ptychographic reconstruction as indicated by a line-out (**c**) and a magnified view (**d**) of two areas. The blurring in ptychography is due to continuous dendrite dissolution as the aperture scans over the field of view.

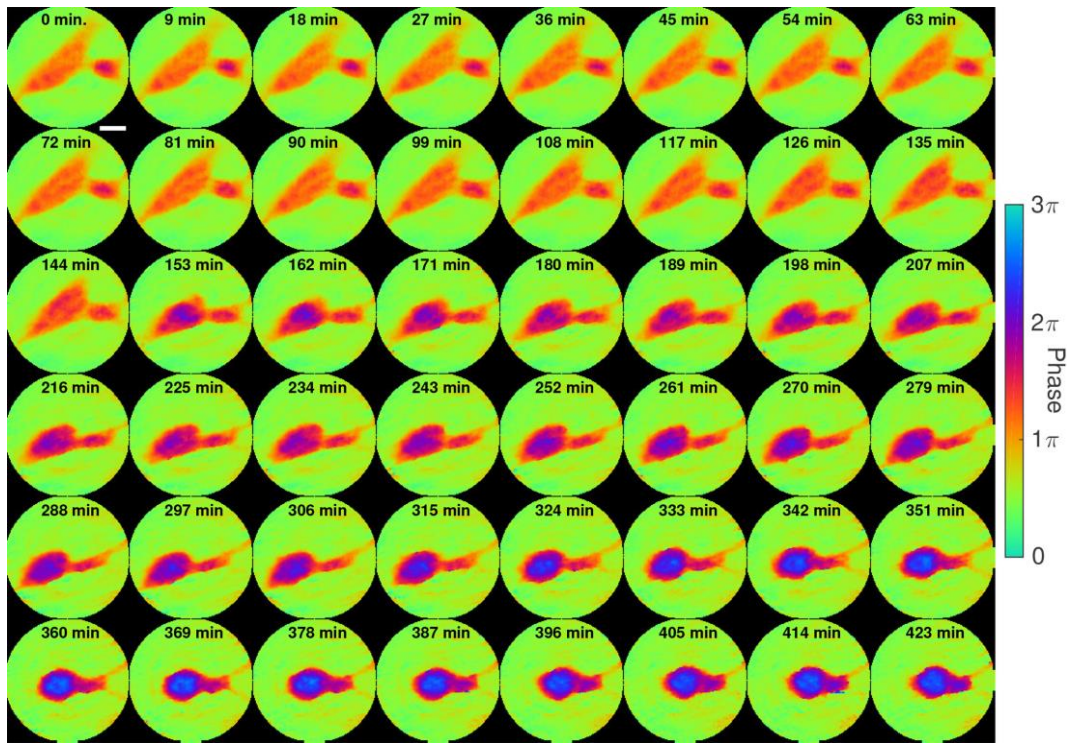

**Figure 4**. **Proof-of-principle experiment on *in situ* CDI with a biological sample**. Phase images of the fusion of glioblastoma cells reconstructed by *in situ* CDI. A smaller cell on the right approached a large cell and initiated cell attachment during the first 144 minutes. Upon attachment, the large cell underwent rapid morphology change and moved left, but the small cell anchored the large cell with thin pseudopodium on the right side of the field of view and began fusing until the 342[th] minute. The cells showed no motility post-fusion, suggesting the occurrence of apoptosis following fusogenic event. Scale bar, 20 μm.

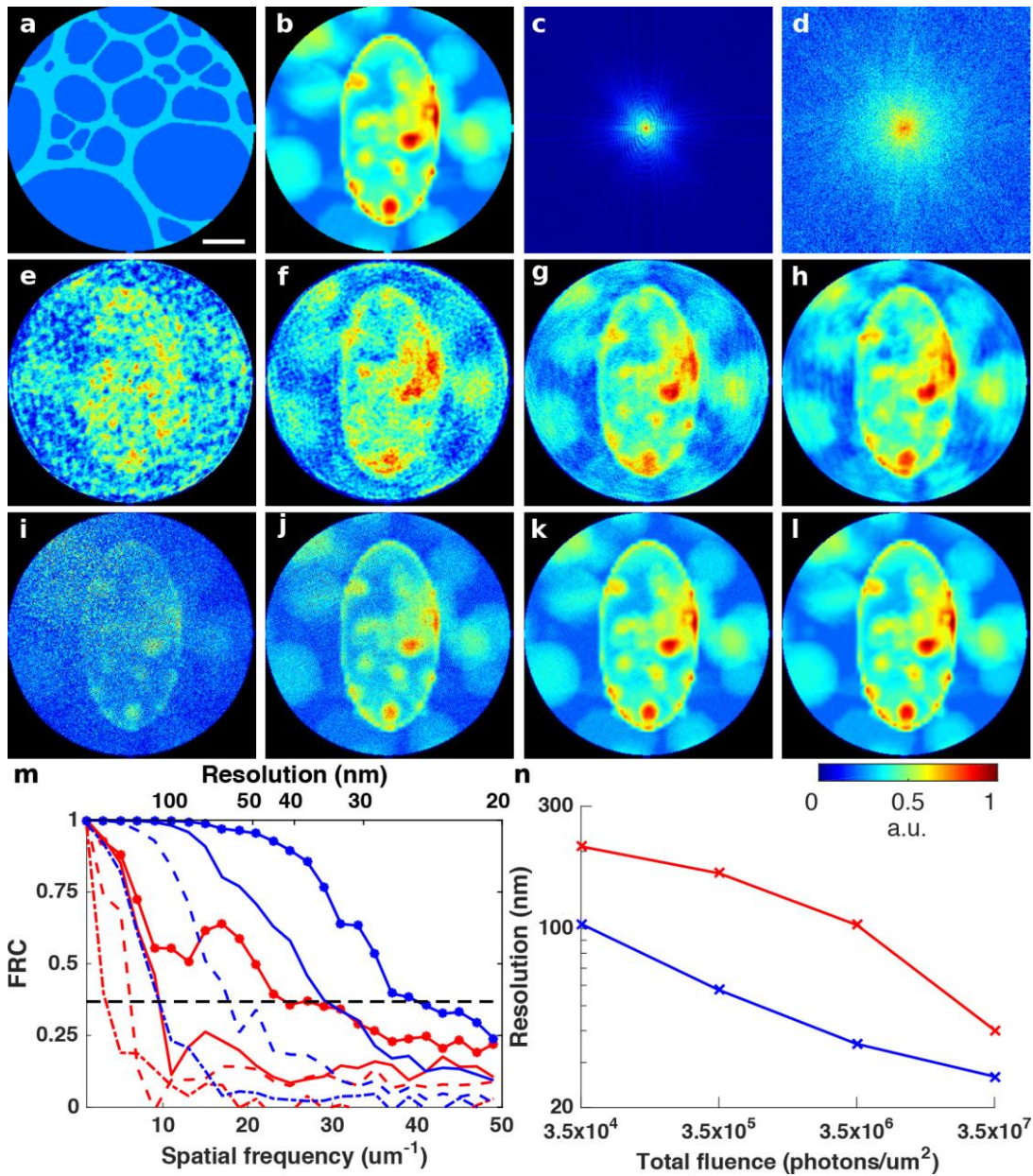

**Figure 5. Numerical simulations of potential significant dose reduction using *in situ* CDI**. **a**, A simulated 20-nm-thick Au pattern in 1 μm $H_2O$ is used as a static structure, in which the diameter of the pinhole is 3 μm. **b**, A simulated biological sample consists of a vesicle and protein aggregates in 1-μm-thick $H_2O$ and masked by a 3 μm pinhole. **c**, Soft X-ray diffraction pattern calculated from the biological sample with a photon energy of 530 eV and a flux of $3.2\times10^7$ photons/μm². Poisson noise was added to the diffraction intensity. **d**, Soft X-ray diffraction pattern calculated from the biological sample with a fluence of $3.5\times10^7$ photons/μm² and the static structure with a fluence of $1.4\times10^{10}$ photons/μm². Poisson noise was added to the diffraction intensity. The center-to-center distance between the biological sample and static structure is 3.8 μm. **e-h**, Image reconstructions of the biological sample without the static structure, with a fluence of $3.5\times10^4$, $3.5\times10^5$, $3.5\times10^6$ and $3.5\times10^7$ photons/μm², respectively. **i-l**, Image reconstructions with the same flux on the biological sample as (**e-h**), but with additional $1.4\times10^{10}$ photons/μm² on the static structure. **m**, Fourier ring correlation of the reconstructions and the model. Red lines correspond to (**e-h**) (dash-dot, dashed, solid,

solid-dot lines, respectively), and blue lines to (**i-l**). **n**, Full-period resolution of each reconstruction determined by the 1/e threshold in the Fourier ring correlation. Scale bar: 400 nm.

## METHODS

**Numerical simulations of *in situ* CDI with coherent X-rays.** To generate a time-evolving structure model for the simulation, we scaled an optical microscopy video of Pb dendrites in an electrochemical cell (Fig. 2c). The thickest part of the Pb dendrites is 500 nm and the thickness of the water layer is 1 μm. Using the complex atomic scattering factor of Pb and $H_2O$ at 8 keV[65], we calculated the projected complex electron density of the structure model as a function of time, $O_t(\boldsymbol{r})$. Next, we created a dual-pinhole aperture consisting of two 1-μm-diameter holes spaced 1.25 μm apart center-to-center. The dual-pinhole illumination function $P(\boldsymbol{r})$ was calculated by propagating the aperture function to the sample plane with a distance of 10 μm. Small random fluctuation is added to $P(\boldsymbol{r})$ to introduce the effect of imperfect illumination function estimate. The diffraction pattern at frame $t$, $I_t(\boldsymbol{k})$, was collected by a 1024×1024 pixel detector,

$$I_t(\boldsymbol{k}) = I_0 \eta \Delta t \left(\frac{r_e \lambda}{a \sigma_1}\right)^2 \left|\Psi_{D,t}(\boldsymbol{k}) + \Psi_S(\boldsymbol{k})\right|^2 \quad, \qquad (9)$$

where $\Psi_{D,t}(\boldsymbol{k})$ and $\Psi_S(\boldsymbol{k})$ are the structure factors of the dynamic and static functions at frame $t$, respectively, $\Psi_{D,t}(\boldsymbol{k}) + \Psi_S(\boldsymbol{k})$ was calculated by using the FFT as $\mathcal{F}[P(\boldsymbol{r}) \cdot O_t(\boldsymbol{r})]$, $I_0$ is the incident photon flux (=$10^{11}$ ph/μm²/s), $\eta$ is the detector efficiency (=0.8), $\Delta t$ is the acquisition time (=10 ms), $r_e$ is the classical electron radius, $\lambda$ is the wavelength, $a$ is the size of illuminated area (=3 μm) , and $\sigma_1$ is the linear oversampling ratio[52] (=2).

**Experiment setup with a HeNe laser**. Our proof-of-principle experiments used a 543 nm HeNe laser (REO) with a power of 5 mW. A collimated beam with a diameter of 800 μm was directed onto a dual pinhole aperture, which consists of two 100 μm pinholes spaced 100 μm apart from edge to edge. The illumination was incident onto the sample 400 μm downstream of the aperture. A 35 mm objective lens was placed immediately downstream of the sample, and far-field diffraction patterns were measured by a 1340×1300 pixel CCD detector (16 bits, Princeton Instruments) at the lens' back focal plane. In order to increase the dynamic range of the diffraction intensity, three separate exposure times, 100, 1,000 and 10,000 ms, were taken and the diffraction patterns were computationally stitched together without missing centers.

**Electrochemical cell preparation**. A sealed fluid cell was assembled to observe the dynamics of Pb dendrites. With the aid of an optical microscope, two platinum wires (diameter = 50.8 um, 99.95% Alfa Aesar) were immersed in a thin layer of a saturated solution of $Pb(NO_3)_2$ (99.5%, SPI-Chem) in deionized water and were encapsulated between two glass microscope coverslips (22×22×0.13 mm). The two glass slides were epoxied together with the platinum wires exposed for making electrical contact.

**Glioblastoma cell preparation**. The glioblastoma cell line U-87 MG was purchased from ATCC (Manassas, Virginia). Cells were cultured in T75 cell culture flask (Thermo Fisher) in Dulbecco's Modified Eagle Medium (DMEM) (Thermo Fisher) with 10% fetal bovine serum (Corning Inc.) and 100 U/ml penicillin-steptomycin (Thermo Fisher) in 37ºC and 5% $CO_2$ incubator. To seed the cells onto coverslips, the cells were treated with TrypLE (Thermo Fisher) for 5 minutes in a 37ºC incubator. The reaction was stopped by adding an equal volume of the complete culture medium. Around 1 million cells were seeded in a 100 mm glass plate with 4 pieces of coverslips inside the plate to allow attachment.

**Potential significant radiation dose reduction using *in situ* CDI**. The dose reduction simulation used 530 eV X-rays to minimize water background and to ensure good cellular contrast. A dual-pinhole aperture with 3 μm diameter pinholes spaced 4 μm apart was used to illuminate the static structure and a biological sample covering 7×7 μm field of view. A 1100×1100 pixel detector with 10 μm detector pixel size was placed 5 cm downstream of sample, with maximum half-period resolution at detector edge of 10.6 nm and an oversampling ratio ($\sigma_1$) of ~2 ($\sigma_1$~4 for single pinhole case). The simulated biological specimen consists of a 2 μm long organelle and various cytosolic components in a 3×3 μm$^2$ region of a 1 μm thick cell. The complex electron density of the biological specimen is calculated using the average composition of protein ($H_{50}C_{30}N_9O_{10}S_1$). Adjacent to the specimen is the static structure, composed of 20 nm thick Au pattern resembling a lacey carbon morphology. The recorded diffraction intensity $I(\boldsymbol{k})$ with 1 s exposure ($\Delta t$) was calculated as

$$I(\boldsymbol{k}) = \eta \Delta t \left(\frac{r_e \lambda}{a \sigma_1}\right)^2 \{I_D |\Psi_D(\boldsymbol{k})|^2 + I_S |\Psi_S(\boldsymbol{k})|^2 + \sqrt{I_D I_S}(\Psi_D(k)\Psi_S^*(k) + \Psi_D^*(k)\Psi_S(k))\} \quad (10)$$

Where $\Psi_D(\boldsymbol{k})$ and $\Psi_S(\boldsymbol{k})$ are the complex waves of the biological specimen and static structure, respectively, calculated using tabulated atomic scattering factors of their respective materials. $I_S$ and $I_D$ are photon fluxes on the static structure and biological specimen, respectively. Eq. (10) is an expansion of Eq. (9) to allow for differential flux through each structure. In the case that $I_D$ equals $I_S$, Eq. (10) is reduced to Eq. (9).

Phase retrieval on the simulated noisy diffraction intensity was performed using OSS[46] with 500 iterations. The reconstruction with the lowest Fourier R-factor in 10 independent runs was used as the final result. Resolution was quantified by the Fourier ring correlation (*FRC*),

$$FRC(\boldsymbol{k}) = \frac{\sum \Psi_m(\boldsymbol{k}) \cdot \Psi_g(\boldsymbol{k})}{\sqrt{\sum |\Psi_m(\boldsymbol{k})|^2 \cdot \sum |\Psi_g(\boldsymbol{k})|^2}} \tag{11}$$

where $\Psi_m(\boldsymbol{k})$ and $\Psi_g(\boldsymbol{k})$ are the complex structure factors of the model and reconstruction, respectively.

**Quantification of radiation dose in simulation.** In the simulation, we estimated the radiation doses ($D$) imparted on the biological specimen as[55,56],

$$D = \left(\frac{P_t}{A}\right)\left(\frac{\mu E}{\rho}\right) \tag{12}$$

where total incident X-ray photons ($P_t$) per unit area ($A$) through a 3 μm pinhole ($P_t/A$) varies from $3.5 \times 10^4$ to $3.5 \times 10^7$ photons/μm². The cell density ($\rho$) is 1.35 g/cm³, and the linear absorption coefficient ($\mu$) of average protein at 530 eV photon energy ($E$) is $1.25 \times 10^4$ cm⁻¹, which gives a mass absorption coefficient ($\mu/\rho$) of $9.26 \times 10^3$ cm²/g. Thus, the total dose delivered to the biological specimen ranges from $2.75 \times 10^3$ to $2.75 \times 10^6$ Gy.